\documentclass[titlepage]{csetr}

\usepackage{graphicx}
\usepackage{subfigure}
\usepackage{url}
\usepackage{mathtools}
\usepackage{underscore}
\usepackage{soul}
\usepackage{makeidx}
\usepackage{cite}
\usepackage{multirow}
\usepackage{colortbl}

\renewcommand{\and}{\hspace{.5cm}}

\trnumber{201304}
\trdate{January 2013}

\title{%
  Providing Trustworthy Contributions via a Reputation Framework in Social Participatory Sensing Systems
}
\author{%
  Haleh Amintoosi\and %
  Salil S. Kanhere
  \\[2em]
  University of New South Wales, Australia \\%
  \email{\{haleha,salilk\}@cse.unsw.edu.au}\\
  \\[3cm]
}

\date{}
\begin{document}
\maketitle

\begin{abstract}
 Social participatory sensing is a newly proposed paradigm that tries to address the limitations of participatory sensing by leveraging online social networks as an infrastructure. A critical issue in the success of this paradigm is to assure the trustworthiness of contributions provided by participants. In this paper, we propose an application-agnostic reputation framework for social participatory sensing systems. Our framework considers both the quality of contribution and the trustworthiness level of participant within the social network. These two aspects are then combined via a fuzzy inference system to arrive at a final trust rating for a contribution. A reputation score is also calculated for each participant as a resultant of the trust ratings assigned to him. We adopt the utilization of PageRank algorithm as the building block for our reputation module. Extensive simulations demonstrate the efficacy of our framework in achieving high overall trust and assigning accurate reputation scores.
\end{abstract}

\section{Introduction}
\label{intro}
Recent advances in mobile technologies have paved the way for a novel paradigm for achieving large-scale city-wide sensing known as Participatory Sensing \cite{Burke}. In Participatory sensing, the key idea is to recruit ordinary people to contribute in sensor data collection using their mobile phones. This revolutionary paradigm has been operationally used to crowdsource information ranging from personal health \cite{DietSense} and prices of consumer goods \cite{PetrolWatch} to environment monitoring \cite{EarPhone}.\\
As a crowdsensing platform, a key challenge in the success of participatory sensing is the recruitment of sufficient participants. Typically, participatory sensing campaigns rely on voluntary contributors without any explicit incentives for participation. The lack of adequate motivation may result in few participants which in turn, reduces the data reliability. Another challenge is the suitability of participants particularly for those tasks which require domain-specific knowledge or expertise \cite{reddyrecruitment}.\\
To address the aforementioned challenges, one proposed idea is to employ online social networks as the underlying substrate for recruiting well-suited contributors \cite{integrate, integratemain}. This marriage of participatory sensing and online social networks, referred to as \emph{social participatory sensing}, offers the following advantages. First, the identification of suitable participants can be done easily through the public profile information such as interests, expertise and education. Second, social ties can act as an effective motivation to contribute to tasks created by friends, since people normally like to be helpful to their friends. Third, it is possible to offer incentives in the form of e-coins \cite{ecash} or reputation points which can be published in participants' profile and seen by others. A real-world instantiation of social participatory sensing was recently presented in \cite{twitter}, wherein, Twitter was used as the underlying social network substrate. The proposed system was tested in the context of two applications: weather radar and noise mapping. Their experiment resulted in a considerable smartphone-based participation from Twitter members even without an incentive structure. This clearly demonstrated the suitability of online social networks as a publish-subscribe infrastructure for tasking/utilizing smartphones and pave the way for ubiquitous crowd-sourced sensing and social collaboration applications.

The open nature of participatory sensing which allows everyone to contribute, while valuable for encouraging participants, facilitates erroneous and untrusted data preparation. When combined with social network, new trust issues arise. For instance, following the devastation incurred due to Hurricane Sandy in the US in October 2012, social media was flooded with misinformation and fake photos \footnote{http://news.yahoo.com/10-fake-photos-hurricane-sandy-075500934.html}. While some of these were easy to identify as fake data (e.g., photoshopped images of sharks swimming in New York streets), several other fake pictures and reports were initially thought to be true.
In fact, the widespread use of social networks, along with fast and easy-to-use dissemination facilities such as re-sharing (a fake photo) or re-tweeting (a false event) make it difficult to identify the origin of the data and investigate its credibility. This clearly highlights the need for a trust system which is responsible for performing necessary validations both from the perspective of data trustworthiness and also the reliability of data contributors. In other words, it is important to know who and with what level of social trustworthiness produces the data and how much of the data can be trusted. While there exist works that address the issue of data trustworthiness in participatory sensing (see Section \ref{rel}), they do not provide means to include social trust and as such cannot be readily adopted for social participatory sensing.\\
In this paper, we present an application agnostic framework to evaluate trust in social participatory sensing systems. Our system independently assesses the quality of the data and the trustworthiness of the participants and combines these metrics using fuzzy logic to arrive at a comprehensive trust rating for each contribution. These trust ratings are then used to calculate and update the reputation score of participants. By adopting a fuzzy approach, our proposed system is able to concretely quantify uncertain and imprecise information, such as trust, which is normally expressed by linguistic terms rather than numerical values. We undertake extensive simulations to demonstrate the effectiveness of our reputation framework and benchmark against the state-of-the-art. The results demonstrate that considering social relations makes trust evaluation more realistic, as it resembles human behaviour in establishing trustful social communications. We also show that our framework is able to quickly adapt to rapid changes in the participant's behaviour by prompt and correct detection and revocation of unreliable contributions and accurate update of participant's reputation score. Moreover, we find that leveraging fuzzy logic provides considerable flexibility in combining the underlying components which leads to better assessment of the trustworthiness of contributions. Our framework results in a considerable increase in the overall trust over a method which solely associates trust based on the quality of contribution.

The rest of the paper is organised as follows. Related work is discussed in Section \ref{rel}. We present the details of our framework in Section \ref{pro}. Simulation results are discussed in Section \ref{sim}. Finally, Section \ref{con} concludes the paper.

\begin{figure}
\centering
\includegraphics[width=12cm]{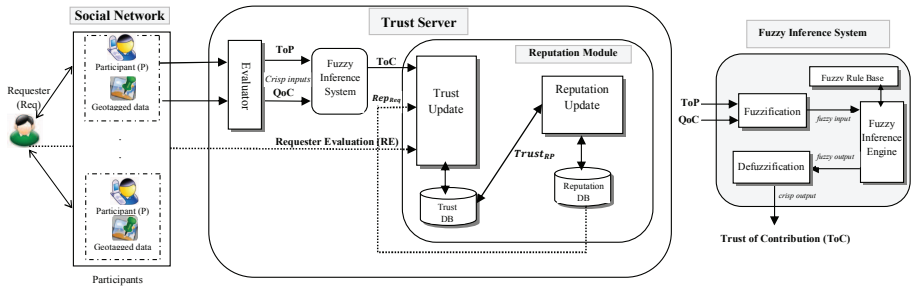}
\caption{Reputation framework architecture}
\label{fig:frm}
\end{figure}

\section{Related Work}
\label{rel}
To the best of our knowledge, the issue of trust in social participatory sensing hasn't been addressed in prior work. As such, we discuss about related research focussing on trust issues in participatory sensing.\\
In a participatory sensing system, trustworthiness can be viewed as the quality of the sensed data. In order to ascertain the data trustworthiness, it is highly desirable to ascertain that the sensor data has been captured from the said location and at the said time. \cite{Lenders} has proposed a secure service which allows participants to tag their content with a spatial timestamp indicating its physical location, which is later used by a co-located infrastructure for verification. A similar approach has been proposed in \cite{LocProof}, in the form of a small piece of metadata issued by a wireless infrastructure which offers a timestamped signed location proof. Since these works rely on external infrastructure, they have limited scalability. Moreover, neither approach will work in situations where the infrastructure is not installed. In our proposed framework, we assume that sensor data is tagged with GPS coordinates/system time before being stored in phone memory, which is then used by trust server for verification. Data trustworthiness has been investigated from another point of view which tries to confirm that uploaded data preserves the characteristics of the original sensed data and has not been changed unintentionally or maliciously. In particular, there are several works which make use of Trusted Platform Module (TPM)\cite{TPM}, which is a micro-controller embedded in the mobile device and provides it with hardware-based cryptography as well as secure storage for sensitive credentials. In \cite{Dua}, each device has a trusted hardware element that implements cryptographic algorithms for content protection. \cite{Imasensor} presents two TPM-based design alternatives: the first architecture relies on a piece of trusted code and the second design incorporates trusted computing primitives into sensors to enable them sign their readings. YouProve \cite{YouProve} is another TPM-based system that allows client applications to directly control the fidelity of data they upload and services to verify that the meaning of source data is preserved.
However, TPM chips are yet to be widely adopted in mobile devices. There is also recent work that does not require TPM. \cite{RFSN} proposes a reputation-based framework which makes use of Beta reputation \cite{beta} to assign a reputation score to each sensor node in a wireless sensor network. Beta reputation has simple updating rules as well as facilitates easy integration of ageing. However, it is less aggressive in penalizing users with poor quality contributions. A reputation framework for participatory sensing was proposed in \cite{Brian}. A watchdog module computes a cooperative rating for each device according to its short-term behaviour which acts as input to the reputation module which utilizes Gompertz function \cite{gompertz} to build a long-term reputation score. Their results show an improvement over the non-trust aggregation based approaches and Beta reputation system. However, the parameters related to the participants' social accountability have not been considered. As such, their system cannot be readily used in our context.

\section{Fuzzy Trust Framework}
\label{pro}
In this section, we explain the proposed framework for evaluating trust and reputation in social participatory sensing systems. An overview of the architecture is presented in Section \ref{arc} followed by a detailed discussion of each component in Section \ref{comp}.

\subsection{Framework Architecture}
\label{arc}
Since our framework attempts to mimic how human's perceive trust, we first present a simple illustrative example. Suppose John is a member of an online social network (e.g., Facebook). He has made a profile and has friended several people. John is a vegetarian and is the member of several vegetarian social communities. He is also on a budget and is keen to spend the least possible amount for his weekly groceries. He decides to leverage his social circle to find out the cheapest stores where he can buy vegetarian products. Specifically, he asks his friends or community members to capture geotagged photos of price labels of vegetarian food items when they are out shopping and send these back to him. One of his friends, Alex decides to help out and provides him with several photos of price labels. In order to decide whether to rely on Alex's contributions, John would naturally take into account two aspects: (i) his personal trust perception of Alex, which would depend on various aspects such as the nature of friendship (close vs. distant), Alex's awareness of vegetarian foods, Alex's location, etc and (ii) the quality of Alex's data which would depend on the quality of the pictures, relevance of products, etc. In other words, John in his mind computes a trust rating for Alex's contribution based on these two aspects. Our proposed trust framework provides a means to obtain such trust ratings by mimicking an approach similar to John's perception of trustworthiness in a scalable and automated manner. This trust rating helps John to select trustable contributions and accordingly plan for his weekend shopping.
Moreover, the trust server provides a reputation score for each of the participating friends, according to the trustworthiness of their successive contributions.

\noindent Fig. \ref{fig:frm} illustrates the architecture of the proposed reputation framework. The social network serves as the underlying publish-subscribe substrate for recruiting friends as participants. In fact, the basic participatory sensing procedures (i.e., task distribution and uploading contributions) are performed by utilizing the social network communication primitives. A person wishing to start a participatory sensing campaign acts as a requester and disseminates the tasks to his friends via email, private message or by writing as a post on their profiles (e.g., Facebook wall). Friends transfer their contributions via email or in the form of a message. We can also benefit from group construction facilities in Facebook or community memberships in Google Plus. The contributions received in response to a campaign are transferred (e.g., by using Facebook Graph API\footnote{http://developers.facebook.com/docs/reference/api/}) to a third party trust server, which incorporates the proposed fuzzy inference system and arrives at an objective trust rating for each contribution. This trust rating is used as a criterion to accept the contribution or revoke it, by comparing against a predefined threshold.

At the end of each campaign, a cumulative objective trust rating, referred to as $Trust_{RP}$ is automatically updated for each participant, which denotes the trustworthiness degree of Requester upon the Participant. $Trust_{RP}$ is dependent on the trustworthiness of the contribution that the participant has prepared for the requester.

For certain campaigns, depending on the nature of task, the requester may desire to add a subjective evaluation in order to indicate how much the contribution is compatible with his needs and expectations. In such a case, this subjective rating is combined with the system-computed rating to update $Trust_{RP}$.

At regular intervals, a reputation score is also calculated for each participant, which is a combination of the trust ratings that requesters have assigned to him. This reputation score is further used as a weight for participant's evaluations, ratings or reviews. More details about trust update, subjective rating and reputation calculation are presented in Section \ref{repsec}.

\subsection{Framework Components}
\label{comp}

This section provides a detailed explanation of the framework components. In particular we focus on the trust sever, fuzzy inference system and reputation module.

\subsubsection{Trust Server}
\label{server}
The trust server is responsible for maintaining and evaluating a comprehensive trust rating for each contribution and calculating a reputation score for each participant. As discussed in Section \ref{intro}, there are two aspects that need to be considered: (1) Quality of Contribution (QoC) and (2) Trust of Participant (ToP). The server maintains a trust database, which contains the required information about participants and the history of their past contributions. When a contribution is received by the trust server, the effective parameters that contribute to the two aforementioned components are evaluated by the Evaluator and then combined to arrive at a single quantitative value for each. The two measures serve as inputs for the fuzzy inference system, which computes the trustworthiness of contribution. In the following, we present a brief discussion about the underlying parameters and the evaluation methods.\\

\noindent\underline{Quality of Contribution (QoC)}\\

In participatory sensing, contributions can be of any modality such as images or sounds. The quality of the data is affected not only by fidelity of the embedded sensor but also the sensing action initiated by the participant. The in-built sensors in mobile devices can vary significantly in precision. Moreover, they may not be correctly calibrated or even worse not functioning correctly, thus providing erroneous data. Participants may also use the sensors improperly while collecting data,(e.g., not focussing on the target when capturing images). Moreover, human-as-sensor applications such as weather radar in \cite{twitter} are exposed to variability in the data quality due to subjectivity. For example, what is hot for one person may be comfortable for another. In order to quantify QoC, a group of parameters must be evaluated such as: relevance to the campaign (e.g., groceries in the above example), ability in determining a particular feature (e.g., price tag), fulfilment of task requirements (e.g., specified diet restrictions), etc. There already exists research that has proposed methods for evaluating the quality of data in participatory sensing. Examples include image processing algorithms proposed in \cite{DietSense} and outlier detection \cite{outlier} for sound-based sensing tasks. Rather than reinventing the wheel, our system relies on these state-of-the-art methods for determining the QoC.\\

\noindent\underline{Trust of Participant (ToP)}\\

\begin{figure}
\centering
\includegraphics[width=12cm]{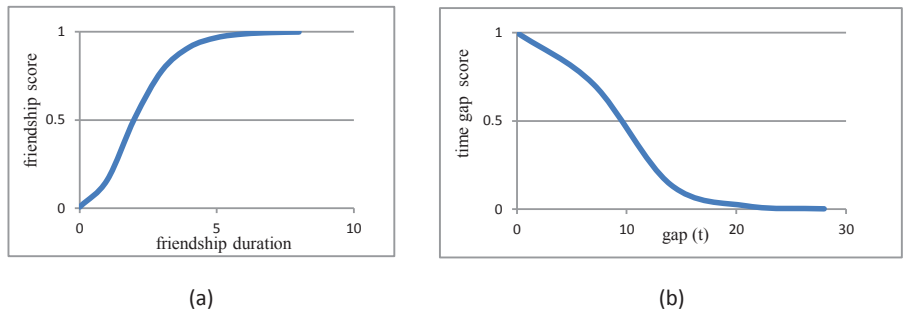}
\caption{(a) Gompertz function for Friendship score \,\,\,\,\, (b) Inverse Gompertz function for time gap score}
\label{fig:gompertz}
\end{figure}

ToP is a combination of personal and social factors. Personal factors consist of the following parameters:\\

\emph{Expertise(E):}
 It is defined as the measure of a participant's knowledge and is particularly important in tasks that require domain expertise. Greater credence is placed in contributions made by a participant who has expertise in the campaign. We propose to use expert finding systems for evaluating expertise. These systems employ social networks analysis and natural language processing (text mining, text classification, and semantic text similarity methods) to analyse explicit information such as public profile data and group memberships as well as implicit information such as textual posts to extract user interests and fields of expertise \cite{expert}. In particular, Dmoz\footnote{http://www.dmoz.org} open directory project is used for expertise classification. Expertise evaluation is done by incorporating text similarity analysis to find a match between the task keywords (e.g., vegetarian) and participant's expertise.
 We assume that the set \emph{TE} contains the Task's required Expertise and \emph{PE} is the set of Participant's Expertises. In this case, the expertise score of each participant is defined as Eq. \ref{eq:eqE}:
 \begin{equation}
   \label{eq:eqE}
   E= \frac{n(TE\cap PE)}{n(TE)}
\end{equation}
where \emph{n(A)} is the number of elements in set A.\\

\emph{Timeliness(T):}
Timeliness measures how promptly a participant performs prescribed tasks. It depends on the contribution response time \emph{(t)} and the task deadline \emph{(d)}. To evaluate this parameter, inverse Gompertz function defined as \begin{math} T(t)= 1 - e^{-be^{-ct}} \end{math} is used because of its compatibility with timeliness evolution: timeliness score is highest when the contribution is received immediately after the task release time. The score begins to decrease as the response time increases, reaching the minimum value when the response is received just before the deadline. In the original inverse Gompertz function, the lower asymptote is zero; it means that the curve approaches to zero in infinity. In our case, timeliness rate will only be zero if contribution is received after the deadline; otherwise, a value between \emph{x} and 1 is assigned to it. It means that the lowest timeliness rating will be \emph{x} if contribution is received before the deadline, and is zero if received after the deadline. So, we modify the function as Eq. \ref{eq:eq1} to calculate the timeliness (T):

 \begin{equation}
  \label{eq:eq1}
  T(t)= \begin{cases}
1 - [(1 - x)e^{-be^{-ct}}] & \text{ if } t<d \\
0& \text{otherwise}
\end{cases}
\end{equation}

\noindent\emph{Locality(L):}
Another significant parameter is locality, which is a measure of the participant's familiarity with the region where the task is to be performed. We argue that contributions received from people with high locality to the tasking region are more trustable than those received from participants who are not local, since the first group is more acquainted with and has better understanding of that region. According to the experimental results presented in \cite{shirazi}, people tend to perform tasks that are near to their home or work place (places that they are considered `local' to them). This implies that if we log the location of participants' contributions, we can estimate their locality. A participant's locality would be highest at locations from where they make maximum number of contributions.
In order to evaluate locality, we assume that the sensing area has been divided to \emph{n} regions, and a vector \emph{V} with the length equal to \emph{n} is defined for each participant, where, \emph{V(i)} is number of samples collected in region \emph{i}. In this case, locality of a participant to region \emph{i} is calculated by Eq. \ref{eq:eq2}:
\begin{equation}
\label{eq:eq2}
 L(i)= \frac{V(i)}{\sum_{i=0}^{n-1} V(i)}
\end{equation}

\begin{table*}
  \centering
  \caption{ Fuzzy rule base for defining ToC according to QoC and ToP}\label{tab:rule}
  \begin{tabular}{|c|c|c|c||c|c|c|c|}\hline
   \scriptsize{Rule no.} & \scriptsize{if QoC} & \scriptsize{and ToP} & \scriptsize{Then ToC} & \scriptsize{Rule no.} & \scriptsize{if QoC} & \scriptsize{and ToP} & \scriptsize{Then ToC} \\\hline
   \scriptsize{1} &  \scriptsize{Low}  & \scriptsize{Low}  &  \scriptsize{VL}   & \scriptsize{9} & \scriptsize{Med2} & \scriptsize{Low}  & \scriptsize{M} \\\hline
   \scriptsize{2} &	\scriptsize{Low}  & \scriptsize{Med1} &  \scriptsize{L}	& \scriptsize{10} &	\scriptsize{Med2} & \scriptsize{Med1} &  \scriptsize{H} \\\hline
   \scriptsize{3} &	\scriptsize{Low}  & \scriptsize{Med2} &  \scriptsize{L}	& \scriptsize{11} &	\scriptsize{Med2} & \scriptsize{Med2} &  \scriptsize{H}\\\hline
   \scriptsize{4} &	\scriptsize{Low}  & \scriptsize{High} &  \scriptsize{M} & \scriptsize{12} & \scriptsize{Med2} & \scriptsize{High} &  \scriptsize{H}\\\hline
   \scriptsize{5} &	\scriptsize{Med1} & \scriptsize{Low} &  \scriptsize{L}	& \scriptsize{14} &	\scriptsize{High} & \scriptsize{Low}&  \scriptsize{H}\\\hline
   \scriptsize{6} &	\scriptsize{Med1} & \scriptsize{Med1} &   \scriptsize{L}	& \scriptsize{14} &	\scriptsize{High} & \scriptsize{Med1}&  \scriptsize{H}\\\hline
   \scriptsize{7} &	\scriptsize{Med1} & \scriptsize{Med2} &  \scriptsize{M}	& \scriptsize{15} &	\scriptsize{High} & \scriptsize{Med2}&  \scriptsize{VH}\\\hline
   \scriptsize{8} &	\scriptsize{Med1} & \scriptsize{High} &  \scriptsize{M}	& \scriptsize{16} &	\scriptsize{High} & \scriptsize{High}&  \scriptsize{VH}\\\hline

  \end{tabular}
\end{table*}

Next, we explain the social factors that affect ToP:\\

\emph{Friendship duration(F):}
In real as well as virtual communications, long lasting friendship relations normally translate to greater trust between two friends. So, friendship duration which is an estimation of friendship length is a prominent parameter in trust development. We use the Gompertz function depicted in Fig. \ref{fig:gompertz}(a) to quantify friendship duration, since its shape is a perfect match for how friendships evolve. Slow growth at start resembles the friendship gestation stage. This is followed by a period of accumulation where the relationship strengthens culminating in a steady stage. As such, the friendship duration is evaluated according to Eq. \ref{eq:eq3}, in which, \emph{b} and \emph{c} are system-defined constants and \emph{t} is the time in years.

\begin{equation}
 \label{eq:eq3}
  F(t)= e^{-be^{-ct}}
  \vspace{-1mm}
  \end{equation}

\noindent\emph{Interaction time gap(I):}
In every friendship relation, interactions happen in form of sending requests and receiving responses. Interaction time gap, measures the time between the consequent interactions and is a good indicator of the strength of friendship ties. If two individuals interact frequently, then it implies that they share a strong relationship, which translates to greater trust.
We propose to use the inverse Gompertz function depicted in Fig. \ref{fig:gompertz}(b) to quantify the interaction time gap, since a smaller time gap indicates stronger relationship, which leads to high social trust and vice-versa.  So, the interaction time gap is evaluated according to Eq. \ref{eq:eq4}, in which, \emph{b} and \emph{c} are system-defined constants and \emph{t} is the gap (in days) between the current time and the Latest Interaction(LI) time.

\begin{equation}
  \label{eq:eq4}
  I(t)= 1-e^{-be^{-ct}}
  \end{equation}

The aforementioned parameters are combined by the Evaluator to arrive at a single value for ToP, as depicted in Eq. \ref{eq:top},
\begin{equation}
\label{eq:top}
ToP= w_1\times E + w_2 \times T + w_3\times L + w_4 \times F + w_5 \times I
\end{equation}
where, \emph{$w_{i}$} is the weight of each parameter, and $\sum_{i=1}^{5}(w_i)$ equals to 1. The adjustment of the weights depends on the nature of the task. For example, in location-based tasks, $w_3$ is set to be considerably high to give more impression to Locality parameter. Similarly, for tasks where real-time information is important, a higher weight may be associated with Timeliness ($w_2$).

\subsubsection{Fuzzy inference system}
\label{fuzzy}
Our proposed framework employs fuzzy logic to calculate a comprehensive trust rating for each contribution, referred to as the Trust of Contribution (ToC). We cover all possible combinations of trust aspects and address them by leveraging fuzzy logic in mimicking the human decision-making process. The inputs to the fuzzy inference system are the crisp values of QoC and ToP. In the following, we describe the fuzzy inference system components.

\begin{figure}[!h]
    \centering
    \subfigure[Membership function for QoC and ToP]
    {
        \includegraphics[height=1in, width=2.5in]{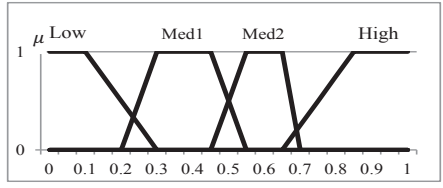}
        \label{fig:mfinput}
    }

    \hspace{0.1cm}
    \subfigure[Membership function for ToC]
    {
        \includegraphics[height=1in,width=2.5in]{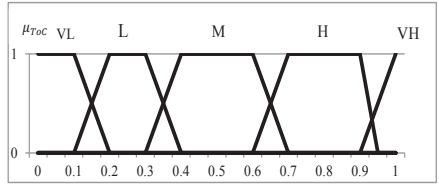}
        \label{fig:mfoutput}
    }
    \caption{Membership functions of input and output linguistic variables}
    \label{fig:mf}
\end{figure}

\noindent\emph{Fuzzifier:}
The fuzzifier converts the crisp values of input parameters into a linguistic variable according to their membership functions.  In other words, it determines the degree to which these inputs belong to each of the corresponding fuzzy sets. The fuzzy sets for QoC, ToP and ToC are defined as: \\ T(QoC)=T(ToP)=\{Low, Med1, Med2, High\}\\
T(ToC)= \{ VL, L, M, H, VH\}.\\
For any set $X$, a membership function on $X$ is any function from $X$ to the real unit interval [0,1]. The membership function which represents a fuzzy set $A$ is usually denoted by $\mu_{A}$. The membership degree $\mu_{A}(x)$ quantifies the grade of membership of the element $x$ to the fuzzy set $A$. The value 0 means that $x$ is not a member of the fuzzy set; the value 1 means that $x$ is fully a member of the fuzzy set. The values between 0 and 1 characterize fuzzy members, which belong to the fuzzy set only partially.\\
Fig.\ref{fig:mfinput} represents the membership function of QoC and ToP and Fig.\ref{fig:mfoutput} depicts the ToC membership function. We used trapezoidal shaped membership functions since they provide adequate representation of the expert knowledge, and at the same time, significantly simplify the process of computation.

\noindent\emph{Inference Engine:}
The role of inference engine is to convert fuzzy inputs (QoC and ToP) to the fuzzy output (ToC) by leveraging If-Then type fuzzy rules. The combination of the above mentioned fuzzy sets create 4*4=16 different states which have been addressed by 16 fuzzy rules as shown in Table \ref{tab:rule}. Fuzzy rules help in describing how we balance the various trust aspects. The rule based design is based on the experience and beliefs on how the system should work. To define the output zone, we used \emph{max-min} composition method as: \begin{math} \mu_{T(ToC)}(ToC)= max[\underset{\substack{X \in T(ToP),\\Y\in T(QoC)}}{min}(\mu_{X}(ToP), \mu_{Y}(QoC))] \end{math}.
The result of the inference engine is the ToC which is a linguistic fuzzy value.

\noindent\emph{Defuzzifier:}
A defuzzifier converts the ToC fuzzy value to a crisp value in the range of [0, 1]. We employed the Centre of Gravity (COG) \cite{cog} defuzzification method, which computes the center of gravity of the area under ToC membership function. COG is perhaps the most commonly used and popular defuzzification technique with the advantage of quick and highly accurate computations.

\subsubsection{Reputation Module}
\label{repsec}
Once the ToC is defined for a contribution, the corresponding requester-participant mutual trust is updated, which is then used to calculate/update the participant's reputation score. In the following, we describe these steps in details:\\
As mentioned before, for some tasks, it is desirable for the requester to assign a subjective rating to participants' contributions. This is particularly relevant for campaigns where it is difficult for the requester to express his real needs, desires or restrictions via task definition. Subjective rating is also useful when the requester does not have enough knowledge about the task and needs an expert review to confirm the validity of the contributions. For example, assume a requester with a strict gluten-free diet who asks his friends to take photos from the price tag and ingredients of gluten-free products. The term gluten-free is generally used to indicate a supposedly harmless level of gluten rather than a complete absence. For those with serious celiac disease, the maximum safe level of gluten in a finished product is even lower than the amount that exists in normal gluten-free products. So, a double check with product ingredients is essential to be performed either by the requester himself or by a nutritionist to assure that it is safe to be consumed.
To be brief, although the objective rating assigned to a contribution is perfect for many tasks, sometimes, a subjective rating is added to reassure the conformance of contribution to the specific needs of requester. In such a case, the need for such subjective evaluation is defined by the requester in the task definition step.

We denote the subjective rating as Requester Evaluation (RE) which implies the trustworthiness of contribution from the requester's point of view. Although RE value can be in any range, in our simulation in Section \ref{sim}, we assume that RE has a value in the range of $(ToC-\mu , ToC+\mu)$, where $\mu$ = 1- $\rho_{Req}$ and $\rho_{Req}$ is the requester's reputation score. For a requester with high reputation score, the value of $\mu$ is small, resulting in RE close to ToC. It means that a requester with high reputation score is likely to assign a rating, which is close to the system-computed rating.

In the absence of subjective ratings, the requester simply relies on the objective ratings assigned by the trust server. In this case, $\mu$ is simply set to zero, resulting in RE=ToC.

Based on the ToC assigned to each contribution, the trust of requester upon the corresponding participant ($Trust_{RP}$) is updated. In fact, we adopt a reward/penalty policy for this update. A participant with ToC values greater than a predefined threshold1$(Th_1)$  is rewarded, and the amount of $\left |ToC-\rho_{Req}*RE \right |$ is added to $Trust_{RP}$. Similarly, a participant with ToC less than a predefined threshold2$(Th_2)$ is penalized, and the amount of $\left |ToC-\rho_{Req}*RE \right |$ is reduced from $Trust_{RP}$. This can be summarized in Eq. \ref{eq:tr}. In our simulations in Section \ref{sim}, we set $(Th_1)=0.7$ and $(Th_2)=0.3$.

\begin{equation}
\label{eq:tr}
\small{Trust_{RP} =\left\{\begin{matrix}
Trust_{RP}+ \left |ToC-\rho_{Req}*RE \right |  & \; if\, ToC>Th_1\\
Trust_{RP}- \left |ToC-\rho_{Req}*RE \right |  & \; if\, ToC<Th_2
\end{matrix}\right.}
\end{equation}
Note that in the this equation, we use the requester's reputation score $(\rho_{Req})$ as a weight for his evaluation($RE$), since we believe an evaluation from a requester with high reputation score is more trustworthy than an evaluation from a low reputable requester.

\begin{figure}
\centering
\includegraphics[width=5cm]{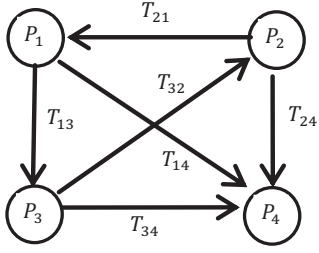}
\caption{A sample social graph of 4 members with mutual trust ratings}
\label{fig:graph}
\end{figure}

This process is repeated for all participants at the end of each sensing campaign, and $Trust_{RP}$ is updated for all of them.

After every $n$ campaigns, $Trust_{RP}$ values upon each active participant act as inputs for reputation module, which updates the participant's reputation score accordingly.

While there are already different crowdsourcing applications of online reputation systems \cite{survey} such as eBay \footnote{http://www.ebay.com/}, Epinions \footnote{http://www.epinions.com/} and Amazon \footnote{http://www.amazon.com/}, we use Web Page ranking algorithms as the basis for computing reputation scores. We draw parallels between the rank of a page in a set of web pages and the reputation score of a member in a social network. Moreover, the weights of links from different pages to a specific page are considered to be equivalent to the trust ratings of one member as determined by the other members of the social network.

Having a set of objects, a ranking algorithm calculates a relative importance of all objects in the set and makes an ordered list according to the importance. Web page ranking algorithms such as PageRank \cite{PageRank} calculate and assign a rank to a web page by analysing the web graph. Roughly speaking, PageRank ranks a page according to how many other pages are pointing at it. This can be described as a reputation system, because the collection of hyperlinks to a given page can be seen as public information that can be combined to derive a reputation score. A single hyperlink to a given web page can be seen as a trust rating of that web page.

In PageRank, the rank of page $P$, denoted by $\rho(P)$ is defined as:
\begin{math} \rho(P)=\frac{\sum\limits_{P_i\rightarrow P}(\rho(P_i))}{L(P_i)} \end{math}
in which, $P_i$ is the set of all pages which have an outgoing link to page $P$, and $L(P_i)$ is the number of outgoing links from page $P_i$.\\

In the original PageRank algorithm, it is assumed that all the outgoing links have equal weights. This is not always true, since not all outgoing links from a web page are equally important. So, we adopted the extension offered in \cite{kaltix} which modifies the above equation as Eq. \ref{eq:pr},
\begin{equation}
\label{eq:pr}
\rho(P)=\sum\limits_{P_i\rightarrow P}{\frac{w_i}{\sum\limits_{P_i\rightarrow P_j}{w_j}} \rho(P_i)}
\end{equation}
in which, $w_i$ is the weight of the outgoing link, and the sum of weights of outgoing links is equal to 1.\\

We explain this further by presenting an illustrative example. Consider the graph in Fig. \ref{fig:graph} in which, $P_1$, $P_2$, $P_3$ and $P_4$ are the social network members.  Links represent friendship relations with weights equal to the mutual trust between the pairs. In this case, according to Eq. \ref{eq:pr}:
\\
\begin{math} \rho(P_1)= T_{21}\times \rho(P_2)\end{math}\\
\begin{math} \rho(P_2)= T_{32}\times \rho(P_3)\end{math}\\
\begin{math} \rho(P_3)= T_{13}\times \rho(P_1)\end{math}\\
\begin{math} \rho(P_4)= T_{14}\times \rho(P_1) + T_{24}\times \rho(P_2) + T_{34}\times \rho(P_3)\end{math}\\
As can be seen in the above expressions, reputation calculation is an iterative process and continues until convergence is obtained. In our simulation in Section \ref{sim}, we assume that the convergence occurs when $\left |\rho_k(P_i)-\rho_{k-1}(P_i) \right| \leq 10^{-10}$ for all $P_i$.   \\

To summarize, once a campaign is launched, participants begin to send a series of contributions. For each contribution, the Evaluator computes a value for QoC and ToP. These values are fed to fuzzy inference engine which calculates ToC for that contribution. The trust of requester upon each participant($Trust_{RP}$) is updated according to his ToC. The server utilizes $Trust_{RP}$ and $\rho_{Req}$ to update the reputation score of each participant.

\section{Experimental Evaluation}
\label{sim}

This section presents simulation-based evaluation of the proposed trust system. The simulation setup is outlined in Section \ref{setup} and the results are in Section \ref{res}.

\subsection{Simulation Setup}
\label{setup}

To undertake the preliminary evaluations outlined herein, we chose to conduct simulations, since real experiments in social participatory sensing are difficult to organise. Simulations afford a controlled environment where we can carefully vary certain parameters and observe the impact on the system performance. We developed a custom Java simulator for this purpose.
We simulate an online social network where 100 members participate in 5000 campaigns, producing one contribution for each. We assume that each member is connected to all others, similar to a social group; So, there are totally 10000 friendship relations. All members can serve both as requesters who launch sensing campaigns and as participants who contribute data to these sensing campaigns.

In our previous work \cite{Mobi}, we assumed of categorizing participants according to the trade-offs between ToP and QoC. We wanted to observe how accurate the system assigns trust ratings to contributions in case of different ToP and QoC levels. Moreover, we artificially created scenarios where participants begin producing contributions with low QoC, which results in a decrease in ToC. We wanted to see if the system is able to quickly detect this transition and revoke low trustable contributions in an accurate and robust manner.

In this paper, instead of categorizing the participants according to ToP and QoC, we designed the categories according to the trade-offs between personal factors and social factors inside ToP, and simply assumed that QoC has a value in the range of $(ToP-\mu , ToP+\mu)$. In fact, we are going to observe how the system reacts to behavioural changes of participants and how much it is successful to update the reputation scores in case of such fluctuations. As mentioned in Section \ref{comp}, ToP parameters can be divided into two groups: social factors which include Friendship duration and Interaction time gap, and personal factors which include Timeliness, Expertise and Locality. In the real-world, there are often situations where a friend with a high rating of social factors (i.e., a very close friend with whom one has repeated interactions) has a low rating for personal factors for a period of time (i.e., does not have related expertise or does not produce timely contributions). It other words, we may have participants who have high social trust, but low personal trust, and vice versa. We have thus 4 different states based on the combination of different levels of personal and social trusts.

Specifically, we assumed that 60 members (out of 100) belong to Category A whereas the remaining 40 belong to Category B, adding the assumption that category A members have high personal trust, while category B members have low personal trust. We also assume that for each member $P_A$ in category A, all other members score $P_A$ with high social trust, and for each member $P_B$ in category B, all other members score $P_B$ with low social trust.

When $P_A$ serves as requester, other members form two subcategories:\\
A-1: which includes 59 members from category A, excluding $P_A$. They have high personal trust and score $P_A$ with high social trust.\\
A-2: which includes 40 members from category B. They have low personal trust and score $P_A$ with high social trust.

Similarly, when $P_B$ serves as requester, other members form two subcategories:\\
B-1: which includes 60 members from category A. They have high personal trust and score $P_B$ with low social trust.\\
B-2: which includes 39 workers from category B, excluding $P_B$. They have low personal trust and score $P_B$ with low social trust.

It is but natural that not all friends in a social network would contribute data to sensing campaigns. As such, we assume that 10\% of the members in category A and 50\% of the members in category B do not upload any data. The rationale for assuming unequal percentages is that the first group are close friends and hence a higher percentage would be willing to contribute, whereas the second group are not so and have less willingness to contribute.

Whenever a task is launched, one of the participants is selected to be the requester. Without loss of generality we assume that tasks are launched in sequential order by the social network members, i.e., member 1 launches the first campaign, member 2 launches the second campaign and so on.\\

\noindent\underline{ToP Parameter Settings}

\begin{table}
  \centering
 \scriptsize{ \caption{ ToP parameter settings}\label{tab:setting}
  \begin{tabular}{|c|c|c|c|}\hline
  \multicolumn{2}{|c}{category A}&\multicolumn{2}{|c|}{category B}\\\hline
  \cline{1-4}
   param&value&param&value\\\hline
   n(PE)&4&n(PE)&2\\\hline
   rt&$\left\{\begin{matrix}
    (0,1]&prob=0.4\\
    (1,7/2]&prob=0.65\\
    (7/2,7]&prob=0.9
    \end{matrix}\right.$&rt&$\left\{\begin{matrix}
(0,1]&prob=0.1\\
(1,7/2]&prob=0.3\\
(7/2,7]&prob=0.5
\end{matrix}\right.$\\\hline
$N_1$&random(0,1)&$N_1$&random(0,0.5)\\\hline
$N_2$&random(0,0.9)&$N_2$&0\\\hline
$N_3$&random(0,0.8)&$N_3$&0\\\hline
t&rand[4,5]&t&rand[0,1]\\\hline
LI&$\left\{\begin{matrix}
    (0,d]&prob=0.8\\
    0&prob=0.2
    \end{matrix}\right.$&LI&$\left\{\begin{matrix}
(0,d]&prob=0.2\\
0&prob=0.8
\end{matrix}\right.$\\\hline
\end{tabular}
}
\end{table}

In the following, we will discuss the initialisation of the various parameters introduced in Section \ref{comp}.

In order to set the Expertise value for a participant, we assume that there are a total of 6 expertise areas defined and that each task needs at most 3 expertise areas $(n(TE)=3)$. To calculate the Expertise score for each participant, we assign a value to $n(PE)$ based on his category, as shown in Table \ref{tab:setting}. The expertise score $E$ is then calculated using Eq. \ref{eq:eqE}.

For Timeliness, we first set the response time \emph{(rt)} for each participant. Considering the task deadline to be 7 days (d=7), \emph{rt} is assigned a value according to the participant's personal category, as depicted in Table \ref{tab:setting}. For example, for a participant $P_A$ belonging to category A, with probability of 0.4, \emph{rt} is at most one day, with the probability=0.65, \emph{rt} is at most half of a week, and with the probability of 0.9, \emph{rt} is at most one week (Note that the greatest probability is 0.9, since with the probability of 0.1 (10\%), $P_A$ does not attend in sensing campaign). \emph{rt} then acts as the input value for Eq. \ref{eq:eq1} which results in Timeliness score $T$ for participant. Other input parameters for Eq. \ref{eq:eq1} have been set as \emph{x}=0.3, \emph{b}= 6, \emph{c}=0.6, and \emph{d}=7 days.

For Locality, we assume that there are a total of 25 regions and that each participant is local to 3 regions (i.e., Locality score $L$ for these 3 regions is 1). We also assume that when a participant has the maximum Locality score to a region, he has a relatively high locality to its surrounding regions. So, Locality score $L$ is assigned to the surrounding regions up to 3 levels of neighborhood,i.e., $N_1$, $N_2$ and $N_3$, based on participant's category, as shown in Table \ref{tab:setting}.

For Friendship duration, as mentioned in Section \ref{comp}, the input parameter $(t)$ is the time (in years) passed from the beginning of friendship establishment. The initial value of \emph{t} is set according to the participant's category, as shown in Table \ref{tab:setting} and a constant value of 0.02 is added to \emph{t} upon each participation. 
$t$ is then serves as the input value for Eq. \ref{eq:eq3} which computes the Friendship duration score $F$ for the participant.
Other input parameters for Eq. \ref{eq:eq3} have been set as \emph{b}=5 and \emph{c}=1.

Finally, for the Interaction time gap, as mentioned in Section \ref{comp}, the input parameter $t$ is the gap (in days) between the current time and the Latest Interaction($LI$) time. We set $LI$ based on the category of each participant, as shown in Table \ref{tab:setting}, and calculate $t$ accordingly. $t$ is then fed to Eq. \ref{eq:eq4} which calculates the Interaction time gap score $I$ for the participant. Other input parameters for Eq. \ref{eq:eq4} have been set as \emph{b}=10 and \emph{c}=0.2.

Once all of the aforementioned parameters are computed, ToP is calculated by simply averaging them. In other words, we simply assume that $w_i=1/5$ in Eq. \ref{eq:top}. QoC is then assigned a value in a range of $(ToP-\mu , ToP+\mu)$ with $\mu=0.1$.

ToC is then calculated and $Trust_{RP}$ is updated according to Eq. \ref{eq:tr}. At intervals, reputation score is also updated for participants. We set the reputation interval to be after every 100 campaigns ($n$=100).

In the first scenario, we assume that ToPs follow the category settings throughout the entire simulation. In the second scenario, we assume that ToP parameters change for a group of participants which results in a transition from one category to another (details in Section \ref{res}).

As mentioned in Section \ref{arc}, a ToC rating is calculated for each contribution and those with ToC lower than a predefined threshold are revoked from further calculations. The ToCs for the non-revoked contributions are then combined to form an overall trust for that campaign. In other words, \begin{math} Overall Trust = \frac{\sum_{i=1}^{n} ToC}{n} \end{math} in which, \emph{n} is the number of non-revoked contributions. The revocation threshold is set to 0.5. We consider the overall trust as the evaluation metric. The greater the overall trust the better the ability of the system to revoke untrusted contributions. Overall trust has a value in the range of [0, 1]. We also calculate the reputation scores for all participants to see whether they reflect the behaviour of participants in normal and transition settings. Reputation score value is a number in the range of [0, 1] with initial value of 0.5 for each participant.

We compare the performance of our framework against the following systems: (1) Baseline-Rep: which follows the approach in \cite{Brian} by calculating a reputation score for each participant according to the QoC of his successive contributions. This reputation score is used as a weight for QoC. In other words, \begin{math} ToC= \sqrt{Rep* QoC}\end{math} (2) Average: which includes ToP but computes the ToC simply as an average of ToP and QoC (3) Fuzzy: our proposed framework.

\subsection{Simulation Results}
\label{res}

\begin{figure}
    \centering
    \includegraphics[width=8cm]{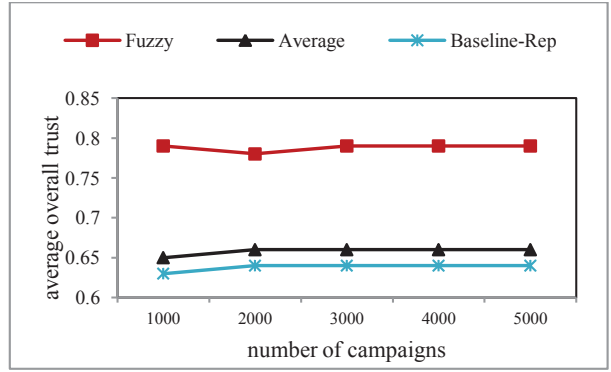}
    \caption{Evolution of average overall trust for all methods, Scenario 1}
    \label{fig:ch1}
\end{figure}

We first present the simulation results for the first scenario. Fig. \ref{fig:ch1} depicts the evolution of the average overall trust as a function of the number of campaigns. As shown in the figure, our fuzzy trust method outperforms the other methods. This confirms its success in mimicking the human trust establishing process by correct settings of fuzzy rules. In particular, we have set the rules in a way that results in early detection and severe punishment of untrusted contributions and also put greater emphasis on highly trusted contributions. The former has been done by assigning a very low (VL) value to ToC in case of low ToP and QoC (i.e., Rule no. 1 in Table \ref{tab:rule}), whereas the latter has been obtained through assigning very high(VH) value to ToC in case of high QoC and above average ToP (i.e., Rule no. 15 and 16 in Table \ref{tab:rule}).

\begin{figure}
    \centering
    \includegraphics[width=8cm]{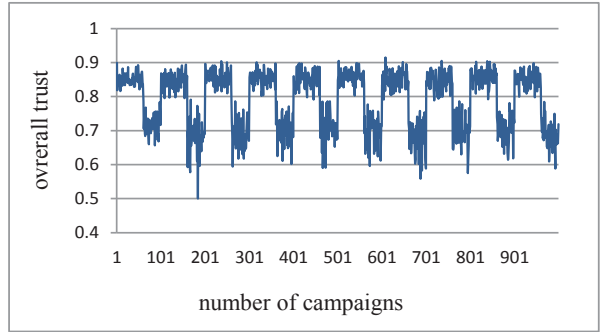}
    \caption{Evolution of overall trust, Fuzzy method, Scenario 1}
    \label{fig:ch2}
\end{figure}
\begin{figure}
    \centering
    \includegraphics[width=8cm]{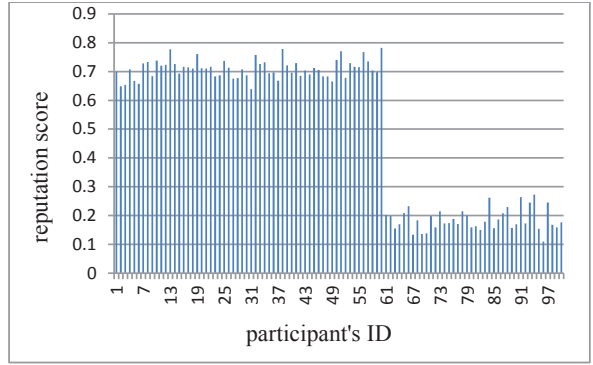}
    \caption{Reputation score for all members, Fuzzy method, Scenario 1}
    \label{fig:ch3}
\end{figure}

Figure \ref{fig:ch2} depicts the evolution of overall trust for 1000 contributions with Fuzzy method. As can be seen in this figure, at each interval containing 100 contributions, two different levels of overall trust are achieved. Remembering the order of requesters which is equal to members' order, higher level of overall trust is obtained when the requester is from category A. So, participants are located either in subcategory A-1 or A-2. This will result in either high ToC values (when participants are from category A-1) or medium ToC values (when participants are from category A-2), which in turn, results in high overall trust. Similarly, lower level of overall trust is obtained when the requester is from category B. So workers are located either in category B-1 or B-2. This will lead to either medium ToC values (when participants are from category B-1) or low ToC values (when participants are from category B-2), which results in low overall trust. This variation is repeated regularly at each interval of 100 contributions.

\begin{figure}
    \centering
    \includegraphics[width=8cm]{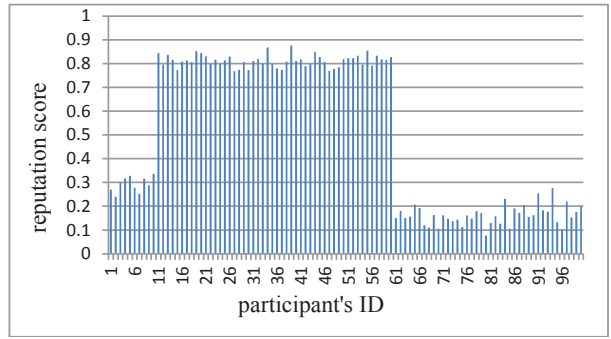}
    \caption{Reputation score for all members at Campaign 4000th, Fuzzy method, Scenario 2}
    \label{fig:ch4}
\end{figure}

\begin{figure}
    \centering
    \includegraphics[width=8cm]{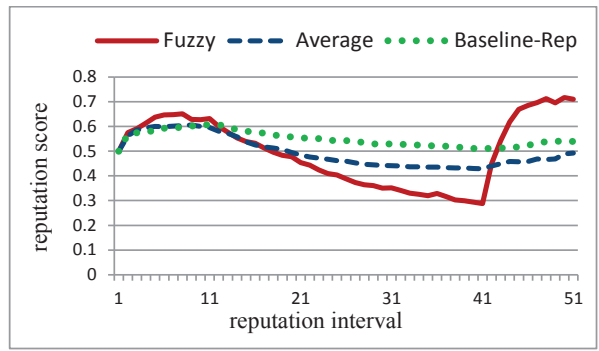}
    \caption{Evolution of Reputation score for participant no.9 in all methods, Scenario 2}
    \label{fig:ch5}
\end{figure}

Fig. \ref{fig:ch3} presents the reputation of 100 participants after attending in 5000 sensing campaigns. As mentioned before, the initial value of reputation score for all participants is 0.5. Category A participants who have high ToPs, produce contributions with high ToC and hence, they get rewarded. This reward results in $trust_{RP}$ increase upon them, which in turn, increases their reputation score. On the contrary, for category B participants with low ToPs, ToCs will also be low, and hence, they are penalized, which results in the reduction of their reputation score. To summarize, our system continually tracks the contributions made over a series of campaigns and detects participants' behaviour, which is accurately reflected in the evolution of the reputation scores.\\
Next, we present results for the second scenario, wherein, the behaviour of participants change for a period of time, which results in a transition from one category to another. This scenario allows us to observe the performance of the schemes in the presence of noise. For example, consider a participant $P_A$ who is in category A, changes his behaviour for a period of time and behaves in a different manner which results in decrease of his personal and (hence) social trust. For example $P_A$ no longer provides timely contributions or does not care enough about the requirements of the task. This behavioural change results in a decrease in his personal trust, and consequently, others score him low with social trust. In other words, a participant may encounter a transition from category A to category B. In this scenario, we assume that 10 from 60 participants of category A transition to category B (e.g., a reduction in their personal and social factor values is created) in the period between 1000\textsuperscript{th} and 4000\textsuperscript{th} campaigns.\\
Fig. \ref{fig:ch4} shows the reputation score of 100 participants at the end of transition period (i.e., after attending in 4000 campaigns). As can be seen in this figure, the reputation of first ten participants who encounter such transition has a considerable decrease in comparison with others not encountering such transition. This again demonstrate the ability of our reputation module to adjust the reputation scores as a reflection of behavioural changes of participants.\\
Finally, Fig. \ref{fig:ch5} shows the reputation score evolution of participant no. 9 encountering such transition between 10\textsuperscript{th} and 40\textsuperscript{th} reputation intervals (between 1000\textsuperscript{th} and 4000\textsuperscript{th} campaigns). As can be observed, our proposed method shows an explicit and considerable reaction to this behavioural change, as compared with other methods. There is a decrease in reputation score due to dishonest behaviour during the transition period. At the end of transition period, transition encountered participant resumes his normal behaviour which results in a considerable increase in his reputation score.

\section{Conclusions}
\label{con}

In this paper, we proposed an application agnostic reputation framework for social participatory sensing system. Our system independently assesses the quality of the data and the trustworthiness of the participants and combines these metrics using fuzzy inference engine to arrive at a comprehensive trust rating for each contribution. The system is then assigns a reputation score to participants by leveraging the concepts utilised in PageRank algorithm. Simulations demonstrated that our scheme increases the overall trust by over 15\% as compared to other methods, and assigns reputation scores to participants in a robust and reliable manner.

\bibliographystyle{plain}

\end{document}